# Environmental effects in the cluster Abell 85 (z=0.055) An HI Imaging Survey and a Dynamical Study

© H. Bravo-Alfaro[1], J.H. van Gorkom[2], C. Caretta[1] & J.M. Islas[1]

[1] Dept. de Astronomia, Guanajuato, Mex.
[2] Dept. of Astronomy, Columbia Univ. NY, USA
[1] Email: hector@astro.ugto.mx

**Abstract:** Based on a VLA HI-imaging through the total volume of the cluster Abell 85, we compare the distribution of HI-rich and HI deficient late type-galaxies with the Intra cluster medium (ICM) drawn by the X-ray emission. Our goals are two fold: first, we study the effects of environment on galaxy evolution as a function of cluster properties, in order to shed light on the morphology transformation of galaxies infalling into clusters. And second, we use the global HI distribution in A85 as an independent method to trace the cluster substructure. We compare this with results of a Δ-test for tracing structures across A85.

**1. Observations:** We carried out a (~90 hrs) volume-complete HI survey on the complex A85/87/89, out to two Abell Radius, with the VLA in its C configuration. The angular resolution is 17"x24" and the covered velocity range is 14,700-18,400 km/s. We compare the HI content throughout the cluster with the intra cluster medium (ICM) drawn by X-ray (XMM). We also compare the HI with the stellar component of individual galaxies seen in optical images taken from the SDSS.

**2. HI Results:** We report 12 HI-detections whose striking distribution around the cluster is shown in Fig. 1: they are asymmetrically clustered, both spatially and in velocity. All but two lie east from A85 and fall within the velocity range 14,900-15,600km/s. All the HI-rich galaxies avoid the densest ICM seen in X-ray (red contours in Fig. 1), but several of them display irregularities in both, the gas and stellar disks. The spatial distribution of the brightest blue galaxies in A85 (Fig. 1), either HI detected (circles) or non-detected (crosses), is very asymmetrical as well. They are defined by taking bj < 17.0, bj-rf < 1.2, within the region (α,δ,vel) of coverage of our VLA survey. The HI-mass upper limit for non-detections is ~5x10$^8$ M$_o$. Most of them appear stripped from their gas component, in spite of being projected on very low density zones of the ICM. On the NE we find the most HI rich objects, all with normal gas content and rotational patterns. On the SE, we detect five galaxies which are likely part of a subcluster infalling to A85, following the X-ray filament, indicated with an ellipse (Durret et al. 2005). Two of the HI detections on the SE are very blue, and show disrupted gas disks and no rotational pattern at all. A clear trend from HI rich to HI deficient (and/or gas disrupted) galaxies is seen as we approach A85 from the SE. Another tendency shows the Hα emission correlated with the distance to A85 (see below).

**3. 3-D search for Substructures:** Analysis of redshifts of 574 galaxies on the A85 region, within the velocity range 0 - 40,000 km/s, confirms the presence of a very complex structure (Fig. 2). The cluster Abell 87, projected 30' SE of A85, is more likely far on the background (v = 40,000 km/s), while the blue galaxies seen on that region (we detect 5 in HI, see Fig. 1) with velocities around 15,000 km/s and projected onto the same region, would rather constitute an infalling group moving towards A85 along the X-ray filament.

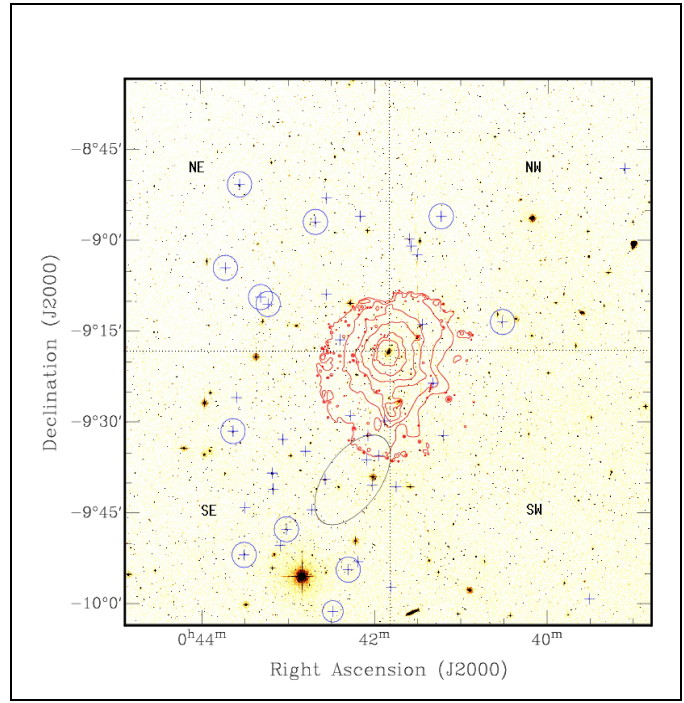

**Fig. 1** Composite plot of A85 showing 12 galaxies detected in HI; positions are shown with encircled. At the center the ICM is drawn by the X-ray (ROSAT) in red contours, with the ellipse sketching the SE filament. Single crosses indicate the distribution of the brightest blue galaxies in A85 which were not detected in HI.

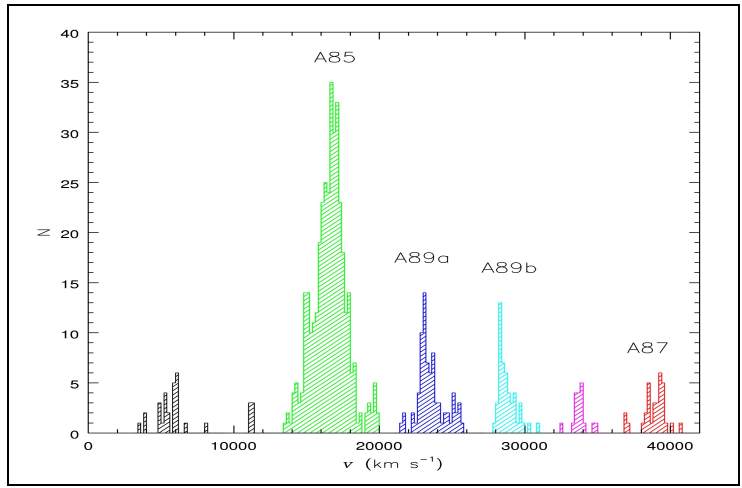

**Fig. 2:** Histogram displaying the distribution of 574 redshifts considering the total range between 0 – 40,000 km/s. The main concentrations are labeled.

Results from a detailed 3-D search for structures (following Dressler and Shectman 1988) are shown in Fig. 3. We consider here only the 367 member galaxies of A85, those within 13,000 km/s and 20,000 km/s (green peak in Fig. 2). In addition to A85, three other structures are unveiled towards the SE: one is slightly south of A85, coincident with the X-ray South Blob (SB, Durret et al. 2005). Two other structures appear farther to the SE, coincident with the X-ray filament (F) and the zone where the blue galaxies are projected (SE).

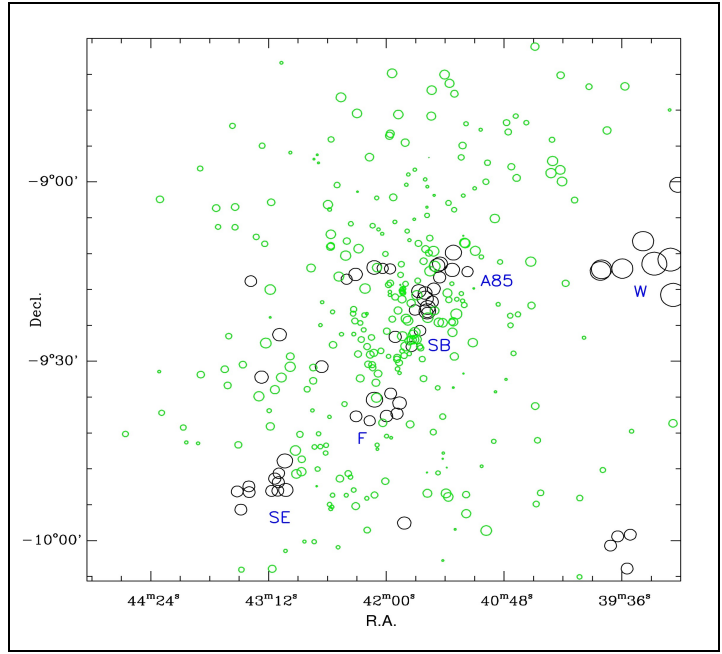

**Fig. 3:** Output from the test to determine substructures in A85. A plot of the Δ-parameter, quantifying the presence of substructure after a 3-D analysis, is shown here. Each circle corresponds to a member of A85, i.e. a galaxy with redshift known within the velocity range of 13,000-20,000 km/s. Larger circles are plotted in black and zones with high concentration of black circles have higher probabilities to constitute groups of galaxies physically linked.

We compare the HI content of galaxies with a preliminary analysis of optical spectral features (Fig. 4). The spectra (from SDSS) of several of these objects indicate different degrees of star formation, going from typical starbursts to dwarf AGN's. In Fig. 4 the intensity of Hα is given by blue circles. This trend where both the HI content and Hα emission diminish when moving to the higher density ICM, seems to confirm that this group of galaxies is currently infalling towards the main core of A85. It remains to understand which of the observed environmental effects are due to the interaction with A85 (ICM or gravitational potential) and which were produced during previous pre-processing events. In the later scenario, both gravitational or ram-pressure like processes could be at work.

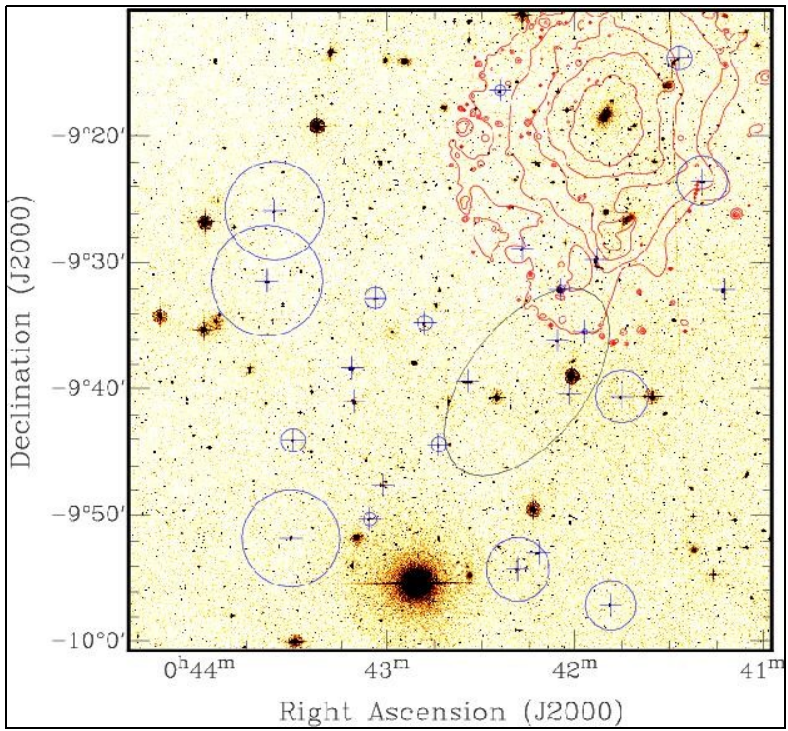

**Fig. 4** Plot of the 29 late type galaxies in the SE of A85. Crosses indicate positions, and circles are proportional to EW of the Hα emission. Red contours draw the main X-ray emission and the ellipse indicates the low brightness filament joining A85 with A87.

**4. Summary:** We confirm the presence of different structures surrounding A85 constituted by groups of galaxies infalling to the main cluster core. The most complex region is the SE, where a series of groups appear moving towards A85 with high relative velocities (at least 1,500 km/s) along the X-ray filament. We report the presence of a conspicuous population of blue galaxies, most of which are gas deficient. We detect in HI only five spirals on that zone but the majority among them display perturbed morphologies, both in their stellar and gaseous disks; furthermore their HI rotational pattern is totally disrupted, indicating strong environmental effects. Most likely these effects are a combination of ram pressure stripping and gravitational effects produced by the global cluster potential. In addition, long distance tidal interactions with neighbor galaxies are certainly at work. The observed trend of Hα emission diminishing as a function of the (projected) cluster distance, give additional support to this scenario. We confirm that galaxies may be strongly disrupted by ram pressure stripping at very large distances from the ICM, as it is proposed by Levy et al. (2007) in clusters such as Perseus I.

### References

1. Dressler, A. & Shectman, S.A. 1988, AJ, 95, 985
2. Durret, F.; Lima Neto, G. B.; Forman, W. 2005, A&A, 432, 809
3. Levy, L., Rose, J. A., van Gorkom, J. H. & Chaboyer, B. 2007, AJ, 133, 1104